\documentclass[aps,pre,showpacs,amsmath,amssymb,superscriptaddress,reprint,10pt]{revtex4-1}

\usepackage{bm}
\usepackage{graphicx}
\usepackage{xcolor}
\usepackage{url}
\usepackage[normalem]{ulem}
\usepackage{times}
\usepackage{bbm}
\usepackage{mathrsfs}
\usepackage{tikz}

\usepackage[colorlinks=true,breaklinks=true,allcolors=blue]{hyperref}
\makeatletter
\let\cat@comma@active\@empty
\makeatother

\newcommand*{\appheading}[1][Appendix]{%
  \setcounter{secnumdepth}{0}\section{#1}\setcounter{secnumdepth}{3}%
  \renewcommand{\thefigure}{\thesection.\arabic{figure}}    
  \numberwithin{equation}{section}
  \numberwithin{figure}{section}
  \setcounter{section}{0}
  \setcounter{equation}{0}
  \setcounter{figure}{0}
}

\newcommand\RR{{\mathbbm{R}}}

\newcommand{\eq}[1]{(\ref{eq:#1})}
\newcommand{\Eq}[1]{Eq.\,\eqref{eq:#1}}

\newcommand{\Fig}[1]{Fig.\,\ref{fig:#1}}
\newcommand{\Figs}[1]{Figs.~\ref{fig:#1}}

\newcommand{\App}[1]{Appendix~\ref{app:#1}}

\newcommand{\sx}{x}
\newcommand{\sy}{y}
\newcommand{\sz}{z}

\allowdisplaybreaks

\begin{document}

\title{Universal equilibrium scaling functions at short times after a quench}  

\author{Markus Karl}
\affiliation{Kirchhoff-Institut f\"ur Physik, Ruprecht-Karls-Universit\"at Heidelberg, Im Neuenheimer Feld 227, 69120 Heidelberg, Germany} 

\author{Halil Cakir} 
\altaffiliation{Present address: Max-Planck-Institut f\"ur Kernphysik, Sau\-pferch\-eckweg 1, 69117 Heidelberg, Germany} 
\affiliation{Kirchhoff-Institut f\"ur Physik, Ruprecht-Karls-Universit\"at Heidelberg, Im Neuenheimer Feld 227, 69120 Heidelberg, Germany} 

\author{Jad C. Halimeh}
\affiliation{Physics Department and Arnold Sommerfeld Center for Theoretical Physics, Ludwig-Maximilians-Universit\"at M\"unchen, D-80333 M\"unchen, Germany}

\author{Markus K. Oberthaler}
\affiliation{Kirchhoff-Institut f\"ur Physik, Ruprecht-Karls-Universit\"at Heidelberg, Im Neuenheimer Feld 227, 69120 Heidelberg, Germany} 

\author{Michael Kastner} 
\affiliation{National Institute for Theoretical Physics (NITheP), Stellenbosch 7600, South Africa} 
\affiliation{Institute of Theoretical Physics, Department of Physics, University of Stellenbosch, Stellenbosch 7600, South Africa}

\author{Thomas Gasenzer}
\email{t.gasenzer@uni-heidelberg.de} 
\affiliation{Kirchhoff-Institut f\"ur Physik, Ruprecht-Karls-Universit\"at Heidelberg, Im Neuenheimer Feld 227, 69120 Heidelberg, Germany} 

\date{\today}

\begin{abstract}
By analyzing spin--spin correlation functions at relatively short distances, we show that equilibrium near-critical properties can be extracted at short times after quenches into the vicinity of a quantum critical point. The time scales after which equilibrium properties can be extracted are sufficiently short so that the proposed scheme should be viable for quantum simulators of spin models based on ultracold atoms or trapped ions. Our results, analytic as well as numeric, are for one-dimensional spin models, either integrable or nonintegrable, but we expect our conclusions to be valid in higher dimensions as well.
\end{abstract}


\maketitle 


The concept of universality asserts that certain properties of a physical system are largely independent of its details. The generality of the concept of universality, and the reasons behind its emergence, have intrigued physicists for decades, and the research efforts culminated in the development of the renormalization group \cite{Kadanoff66,Wilson71,Wilson75}. A prime example are critical phenomena in the vicinity of a continuous phase transition, where only the dimensionalities of space and the order parameter and gross features of the interactions, but not their microscopic details, are relevant. Within a given universality class, this permits a unified description, valid close to the critical point, of systems as diverse as superfluids and magnetic crystals. 

Scaling is closely related to universality, and the two can be seen as twin concepts \cite{Kadanoff90}. For example, universality at a continuous phase transition manifests in the form of power laws in the vicinity of the critical point, where the values of the corresponding exponents are universal quantities. More generally, it implies that the equation of state as well as scaling functions describing certain asymptotic temporal and spatial dependences of correlation and response functions become universal once nonuniversal metric factors have been fixed. Measuring such scaling laws provides therefore a way for experimentally verifying universality, and for extracting the numerical values of critical exponents and critical amplitude ratios that characterize the different universality classes. 

Scaling laws are asymptotic laws in more than one way. Rigorously speaking, to measure critical exponents of an equilibrium (thermal or quantum) phase transition, a system of asymptotically large size is required, in the long-time limit to reach equilibrium, and at parameter values asymptotically close to the critical point. For practical purposes, however, milder conditions are sufficient, as otherwise universality would never have been observed in nature. For sufficiently large systems, after sufficiently long times, there will be some finite region around the phase transition point in which universal laws can be observed to good accuracy. The size of that region is a crucial quantity, as it determines whether or not universal quantities are accessible with the level of control and precision in a given experiment. In the past fifteen years, thermal as well as quantum phase transitions have been investigated in experiments with trapped ultracold atoms \cite{Greiner_etal02,Donner_etal07,Baumann_etal10,Haller_etal10,Zhang_etal12}. The advantage of these experimental platforms is their high level of control and flexibility, allowing the experimentalist to emulate a variety of models, control their parameters, and measure their properties with high accuracy and spatial resolution. In some ultracold-atom experiments, dynamical protocols have been used to probe Kibble-Zurek-type scaling laws by slowly ramping a control parameter across the phase transition point \cite{Navon_etal15,Braun_etal15}. 

Here we explore a unitary dynamical protocol based on sudden quenches into the vicinity of a quantum critical point. We focus on correlation functions and correlation lengths at fairly short times after the quench, and compare them with the corresponding equilibrium quantities. Such a protocol has recently been used in experiments with a two-component Bose gas \cite{Nicklas:2015gwa}. 
The scheme has the advantage of being much less affected by the difficulties near-adiabatic protocols face due to the finite lifetime of the trapped atomic gas samples. 
Universal dynamics and scaling after a quantum quench has been analyzed in different contexts in Refs.~\cite{Lamacraft2007.PhysRevLett.98.160404,Rossini2009a.PhysRevLett.102.127204,Mitra2012a.PRL109z0601M,DallaTorre2013.PhysRevLett.110.090404,Gambassi2011a.EPL95.6,Sciolla2013a.PhysRevB.88.201110,Smacchia2015a.PhysRevB.91.205136,Chiocchetta2015a.PhysRevB.91.220302,Maraga2015a.PhysRevE.92.042151,Maraga2016a.160201763M,Chiocchetta2016a.PhysRevB.94.134311,Chiocchetta:2016waa,Chiocchetta2016b.161202419C,Marino2016a.PhysRevLett.116.070407,Marino2016PhRvB..94h5150M,Damle1996a.PhysRevA.54.5037,Berges:2008wm,Berges:2008sr,Scheppach:2009wu,Nowak:2011sk,Schole:2012kt,Nowak:2012gd,Karl:2013mn,Karl:2013kua,Orioli:2015dxa,Mukerjee2007a.PhysRevB.76.104519,Williamson2016a.PhysRevLett.116.025301,Hofmann2014PhRvL.113i5702H,Williamson2016a.PhysRevA.94.023608,Bourges2016a.arXiv161108922B.PhysRevA.95.023616}, partly motivated by earlier work based on boundary critical phenomena such as critical ageing, coarsening  and phase-ordering kinetics in classical systems \cite{Hohenberg1977a,Janssen1979a,Diehl1986a,Janssen1989a,Janssen1992a,Diehl:1993az,Diehl1993b,Li1995a.PhysRevLett.74.3396,Okano1997a,Calabrese2002a.PhysRevE.65.066120,Calabrese2005a.JPA38.05.R133,Gambassi2006a.JPAConfSer.40.2006.13,Bray1991a.PhysRevLett.67.2670,Bray1994a.AdvPhys.43.357,Bray2000PhRvL..84.1503B}.

Our goal is to identify, in clean models and by means of well-controlled analytic and numeric approaches, the regimes and parameter values for which universal equilibrium scaling functions can be extracted at short (and hence experimentally accessible) times after the quench. The study is conducted for two models with complementary properties, both of which within reach of cold-atom-based quantum simulators. 
(a) The one-dimensional transverse-field Ising model (TFIM), for which analytic results can be obtained; this model is integrable and its states, in the long-time limit, can be described by Generalized Gibbs ensembles \cite{Jaynes1957a,Rigol2007a,Polkovnikov_etal11,Langen:2016vdb}.
(b) The one-dimensional anti-ferromagnetic $\mathrm{XXZ}$ chain in a transverse field; for the chosen parameter values this model is nonintegrable and is expected to thermalize at late times.

Our main result is encouraging: universal properties can be observed for fairly small system sizes of $\mathscr{O}(10^2)$ particles, after short times of only $\mathscr{O}(1/J)$ in the spin--spin coupling $J$.
This regime can be reached in many cold-atom-based experimental platforms, including the one in Ref.~\cite{Nicklas:2015gwa} where such a quench protocol has already been successfully implemented. 
The proposed protocol is therefore expected to be a viable tool for extracting  universal quantities in a variety of experimental settings. 

Our results show that the extraction of equilibrium scaling functions is feasible not only for nonintegrable systems, but also in the integrable case, despite the presence of conserved quantities that prevent thermalization in general. 
These findings are supported by exact computations of correlation functions for the one-dimensional TFIM via a determinant formula, by analytical calculations for the TFIM in the continuum and scaling limits, and by numerical simulations for the one-dimensional $\mathrm{XXZ}$ model using a matrix-product-state-based simulation method for infinite systems.

{\em Quench protocol.---}We consider a family of Hamiltonians, parametrized by $\lambda\in\RR$, that are the sum of two noncommuting terms,
\begin{equation}\label{eq:H_gen}
H(\lambda)=H_1 + \lambda H_2.
\end{equation}
The idea of a quantum quench is to prepare the system in the ground state of $H(\lambda_0)$, and then, starting at time $t=0$, time-evolve that state under $H(\lambda)$ with $\lambda\neq\lambda_0$. 
A quench is rather simple to implement in many ultracold-atom platforms, and it has become a standard tool for experimentally probing nonequilibrium properties of quantum systems. 

The real parameter $\lambda$ in \Eq{H_gen} controls the relevance of $H_2$, and in many cases, at some critical parameter value $\lambda_\text{c}$, a quantum phase transition occurs, i.e., an abrupt change of the ground state properties of $H$ when considered in the thermodynamic limit. 
We introduce the parameter $\varepsilon=\lambda/\lambda_\text{c}-1$ as a measure of the parametric distance from the quantum critical point $\lambda_\text{c}$. To probe universal scaling laws, the system is quenched from $\varepsilon\gg1$ into the vicinity of the quantum phase transition at $\varepsilon=0$.

{\em Transverse-field Ising model.---}We consider the TFIM Hamiltonian
\begin{equation}\label{eq:H_TFIM}
H(h)=-J\sum_{n=1}^N\sigma_n^\sz\sigma_{n+1}^\sz - h\sum_{n=1}^N\sigma_n^\sx,
\end{equation}
describing a chain of $N$ spin-$1/2$ degrees of freedom with periodic boundary conditions. The spins are exposed to an external magnetic field $h\in \RR$ in $\sx$-direction, and $J>0$ is the strength of a ferromagnetic exchange coupling in $\sz$-direction. Without loss of generality we set $J=1$, which fixes the energy scale. At zero temperature and in the thermodynamic limit the model has quantum phase transitions at $|h|=h_\text{c}=1$, separating a ferromagnetic phase at $\lvert h\rvert<1$ from paramagnetic phases at $\lvert h \rvert>1$, with $\langle\sigma_n^\sz\rangle$ being the order parameter of the transition. The Hamiltonian \eq{H_TFIM} can be mapped onto non-interacting fermions, and the spectrum and energy eigenstates can be derived analytically \cite{Pfeuty70}. 

Calculating the dynamics of equal-time spin--spin correlation functions essentially amounts to diagonalizing an $N\times N$ matrix \cite{DerzhkoKrokhmalskii98}. 
Alternatively, for a quench from $h_0$ to $h$, and for an infinite chain, calculating the dynamics of equal-time correlations $C^{\sz\sz}(t,\ell)=\langle\sigma_n^\sz\sigma_{n+\ell}^\sz\rangle(t)$ between sites that are a distance $\ell$ apart can be reduced to computing the determinant of an $\ell\times\ell$ matrix (Eq.~(57) in \cite{CalabreseEsslerFagotti12}). 
We study quenches from $h_0\gg1$, which is equivalent to preparing an essentially $x$-polarized initial state, to a final value $h$ that is at a distance $\varepsilon=h-1$ from the critical value $h_\text{c}=1$.

\begin{figure}\centering
\includegraphics[width=0.99\linewidth]{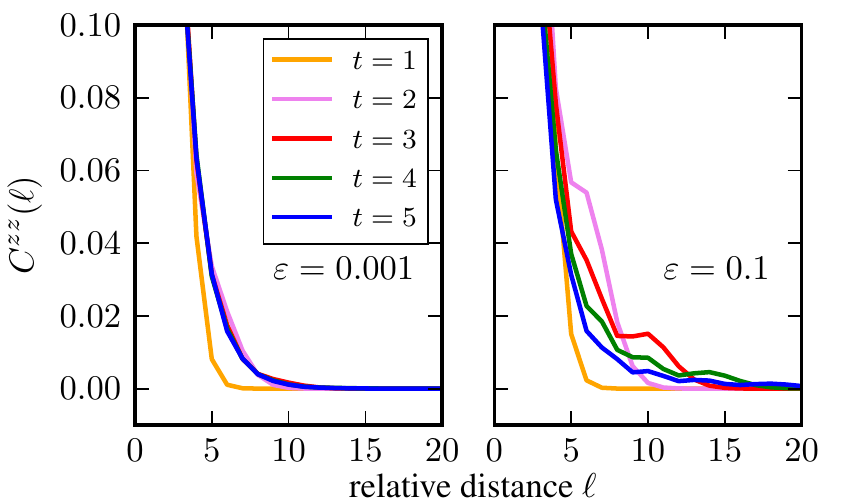}
\caption{\label{fig:Corr_vs_ell}%
Order-parameter correlations $C^{\sz\sz}(t,\ell)$ of the one-dimensional transverse-field Ising model as a function of the spin--spin distance $\ell$, for various fixed times $t$ after a quench from $h_0=\infty$ to $\varepsilon=h-1=0.001$ (left), and $\varepsilon=0.1$ (right).
Time is given in units of $1/J=1$. 
A quasi-stationary short-distance fall-off of the correlations is reached within a few times $1/J$. 
}%
\end{figure}

{\em Quench dynamics.---}For quenches to $\varepsilon\gtrsim0$, i.e., within the paramagnetic phase and into the close vicinity of the quantum critical point, the correlation function $C^{\sz\sz}(\ell)$ shows overdamped dynamics (\Fig{Corr_vs_ell} left).
At least for those distances $\ell$ where $C^{\sz\sz}$ is large enough to be detected experimentally, correlations approach their stationary value already on a time scale of the order of $1$. 
At first, this finding comes as a surprise, as theoretical arguments indicate that, at least for certain types of quenches, full stationarity is reached only at times several orders of magnitude later, even up to $10^{20}$ (cf.~Sec.~2.2.4 of 
\cite{CalabreseEsslerFagotti12}, and the discussion of \Fig{ScaleEll1}  below).
For quenches that end further away from the quantum critical point, like the one in \Fig{Corr_vs_ell} (right) for $\varepsilon=0.1$, oscillations are superimposed on the building-up of correlations. Relaxation time scales are somewhat longer in that case, in the range of a few or even tens of $1/J=1$, but the short-distance decay of $C^{\sz\sz}$ again reaches its stationary values at a time of the order of $1$.

\begin{figure*}
    \centering
    \includegraphics[]{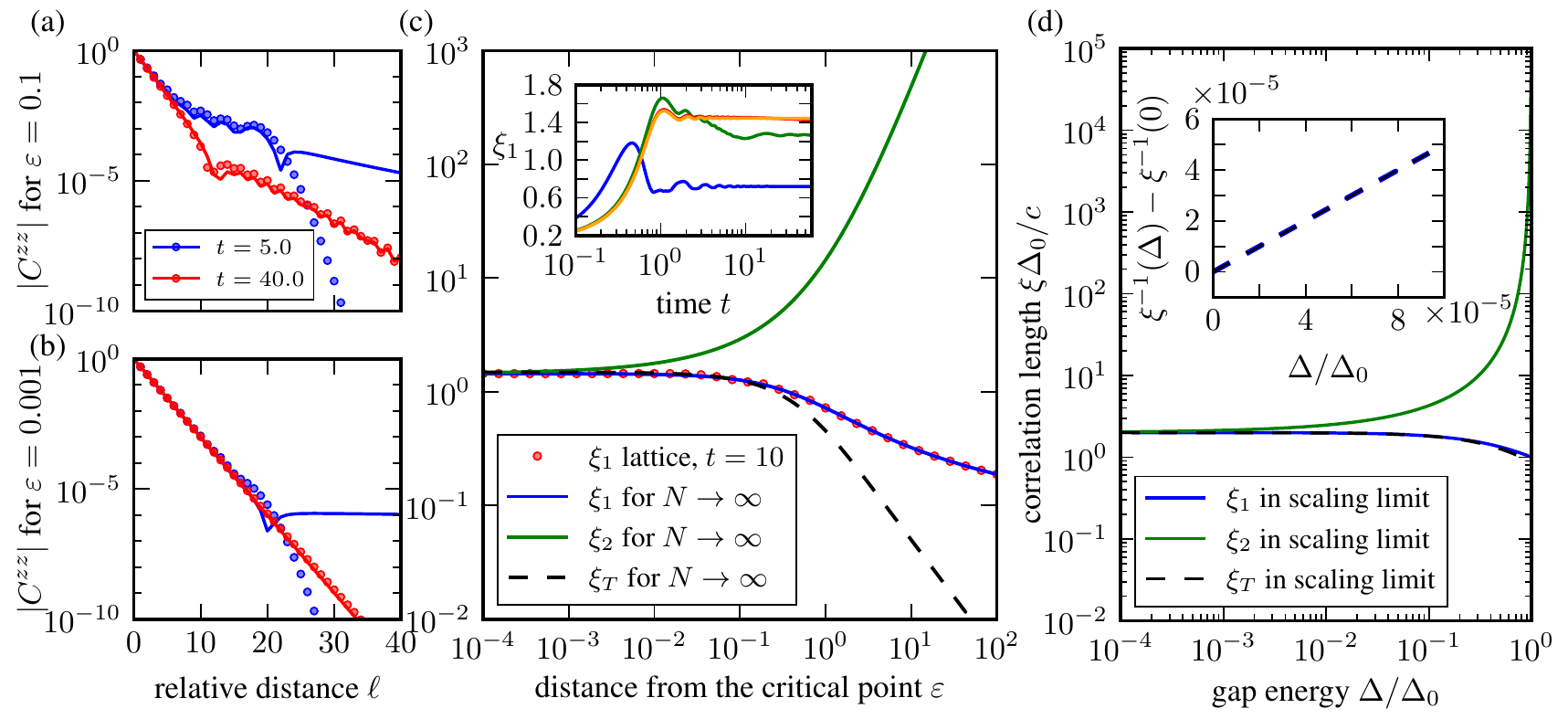}
    \caption{\label{fig:correlation_lengths} 
   Order-parameter correlation functions and correlation lengths after a quench from $h_{0}\gg1$ into the vicinity of the quantum phase transition of the one-dimensional transverse-field Ising model.
   (a) Correlation function $|C^{\sz\sz}|$ as a function of the spin--spin separation $\ell$ at two different times $t$ after the quench from $h_{0}=10^{3}$ to $\varepsilon=h-1=0.1$. 
   Circles mark the exact solution obtained by diagonalisation in terms of Bogoliubov fermions while the solid line shows the approximate function \eq{CzzImproved}.
   Note the logarithmic scale. 
   The envelope of the correlations shows approximately exponential decay, $|C^{\sz\sz}|\sim\exp(-\ell/\xi_{i})$, with different correlation lengths $\xi_{i}$, $i=1,2$, below and above a characteristic scale $\ell_{1}(t)$, see main text.
   (b) As in (a), but for a quench to $\varepsilon=0.001$.
   (c) Inset: Time evolution of $\xi_{1}$, obtained by fitting exponential functions to the short-distance fall-off of $|C^{\sz\sz}|$ for finite spin chains. Data for four different quenches are shown, with final quench parameters $\varepsilon=1.0$ (blue), $0.1$ (green), $0.01$ (red), and $0.001$ (yellow). 
   Main graph: 
   Correlation length $\xi_{1}$ at $t=10$, as a function of the post-quench parameter $\varepsilon$. Red circles 
   are obtained from exponential fits to the data, the blue and green lines are the analytical expressions \eqref{eq:h0oo} for $\xi_{1}$ and \eq{xi2h0oo} for $\xi_{2}$, respectively. 
   $\xi_{T}$ marks the thermal correlation length \eq{xih} for an effective temperature $T=1.58$.
   (d) Analytic results for the transverse-field Ising model in the scaling limit. Main graph: Asymptotic post-quench correlation lengths $\xi_{1,2}(t\to\infty;\Delta)$ (blue and green lines, respectively) as given in Eqs.~\eqref{eq:xionesl} and \eqref{eq:xionesl2}, after a quench from $\Delta_{0}=1$ to $\Delta$. 
   The black dashed line shows the thermal correlation length $\xi_{T}$ at an effective temperature $T=T_{\text{eff}}=2\Delta_{0}/\pi$. 
   This choice of temperature ensures that $\xi_{T_{\text{eff}}}$ and the post-quench correlation lengths $\xi_{1,2}(t)$ coincide at the critical point, i.e., $\xi_{T_{\text{eff}}}=\lim_{t\to\infty}\xi_{1,2}(t)$ for $\Delta=0$. 
   The inset shows that, for $\Delta/\Delta_{0}\ll1$, the same linear corrections apply to $\xi_{1}$ and $\xi_{T_{\text{eff}}}$.
}
\end{figure*}

The rapid convergence to stationarity suggests to use analytic asymptotic expressions, valid at late times $t$ and large distances $\ell$, also for describing the behavior at early times and moderate distances.  
To this purpose, building on a formula originally put forward in \cite{CalabreseEsslerFagotti12}, we propose an improved analytic expression for the spatio-temporal behavior of the order parameter correlations of the TFIM,
\begin{align}\label{eq:CzzImproved}
C&^{\sz\sz}(t,\ell)
 \simeq C_{0}(h_{0},h)e^{-\ell/\xi_{1}(h_{0},h)}
+\left(h^{2}-1\right)^{1/4}\sqrt{4h}
 \nonumber\\
 &\times\ \int_{-\pi}^{\pi}\frac{\mathrm dk}{\pi}
    \left[\frac{n_{\text{BF}}(k)}{1-n_{\text{BF}}(k)}\right]^{1/2}
    \frac{\sin[2\omega_{\text{BF}}(k;h)t-k\ell]}{\omega_{\text{BF}}(k;h)}
 \nonumber\\
 &\ \ \times\ \exp\Big(\int_{0}^{\pi}\frac{\mathrm dk}{\pi}\ln|1-2n_{\text{BF}}(k)|
 \nonumber\\
 &\qquad\times   
 \big\{\ell+\Theta[\ell-2v_{\text{BF}}(k;h)t][2v_{\text{BF}}(k;h)t-\ell]\big\}  \Big).
\end{align}
Here, $\Theta$ denotes the Heaviside step function, and all parameters and functions appearing in \eq{CzzImproved} are specific to a quench from $h_{0}>1$ to $h>1$, and have been derived in  \cite{CalabreseEsslerFagotti12}, specifically, the amplitude
\begin{equation}
   C_{0}(h_{0},h)=\left[\frac{(h_{0}-h)h\sqrt{h_{0}^{2}-1}}{(h_{0}+h)(hh_{0}-1)}\right]^{1/2},
   \quad1<h<h_{0},
\end{equation}
and the inverse correlation length
\begin{multline}\label{eq:xione}
    \xi_1^{-1}(h_{0},h) = \Theta(h-1) \Theta(h_0-1) \ln\min\left\{h_0, h_1\right\}\\
               -\frac{1}{2\pi} \int_{-\pi}^{\pi} \mathrm dk \,\ln\left|1 - 2 n_{\text{BF}}(k)\right|,
\end{multline}
where one defines
\begin{equation}
   h_1 = \frac{1 + h h_0 + \sqrt{\left(h^2-1\right) \left(h_0^2-1\right)}}{h + h_0}.
\end{equation}
The integrands in \Eq{CzzImproved} contain the mode occupation numbers of the Bogoliubov fermions that diagonalize \eq{H_TFIM} after the quench,
\begin{equation}
  \label{eq:OccNoBogF}
  n_{\text{BF}}(k;h_{0},h)
  =\frac{1}{2}-2\frac{h_{0}h+1-(h_{0}+h)\cos(k)}{\omega_{\text{BF}}(k;h)\omega_{\text{BF}}(k;h_{0})},
\end{equation}
with mode frequencies
\begin{equation}
\omega_{\text{BF}}(k;h)=2\sqrt{h^{2}+1-2h\cos(k)}.
\end{equation}
From these, the group velocity is obtained as 
\begin{equation}
  v_{\text{BF}}(k;h)=\frac{d\omega_{\text{BF}}(k;h)}{dk}.
\end{equation}
Our improved formula \eqref{eq:CzzImproved} has the virtue of being consistent with Eq.~(33) of Ref.~\cite{CalabreseEsslerFagotti12} when quenching to values of $h$ far away from the critical point, and with Eq.~(36) of that reference in the limit of quenching close to the critical point. In addition, \Eq{CzzImproved} provides a much improved approximation also within the full range of quench parameters $h$ in between, at early times $t$ and short distances $\ell$.

In \Figs{correlation_lengths}(a) and (b) we compare exact results for the correlation function $C^{\sz\sz}$ to the asymptotic formula \eq{CzzImproved}, finding very good agreement already at rather early times $t$. In particular, at short distances $\ell\lesssim\ell_{1}(t)$ the asymptotic long-time behavior $C^{\sz\sz}(t,\ell)\simeq \exp(-\ell/\xi_{1})$ is already attained to a very good degree. 
The weaker decay at larger distances is described by the second, integral term in \Eq{CzzImproved}, and is characterized by a second correlation length $\xi_{2}$,
see \App{LongRange} for more details.

{\em Correlation length.---}%
The rapid approach to the asymptotic long-time behavior suggests to try and extract equilibrium (i.e., late-time) properties from data at rather early times $t$ and for distances up to $\ell_1(t)$. We extract the correlation length $\xi_{1}$ from the data in \Fig{Corr_vs_ell} by fitting exponential functions to the short-distance decay of $C^{\sz\sz}$. 
The time-dependence of $\xi_{1}$ is shown, for various $\varepsilon$, in \Fig{correlation_lengths}(c) (inset), where stationarity is observed after a rather short time $t\gtrsim1$ for small $\varepsilon$, and at somewhat later times for larger $\varepsilon$. 

The main graph of \Fig{correlation_lengths}(c) shows the correlation length $\xi_1$, at time $t=10$ after the quench as a function of the quench parameter $\varepsilon$ (red points).
Excellent agreement with the asymptotic late-time limit of $\xi_{1}$, obtained from \Eq{xione} as
\begin{equation}\label{eq:h0oo}
\left.\xi_{1}^{-1}\right|_{h_{0}\to\infty}=\ln(2h),
\end{equation}
and shown as a blue line in \Fig{correlation_lengths}(c), is observed. 

{\em Comparison with thermal equilibrium.---}%
Our findings confirm that, for the quenches studied here, equilibrium correlation lengths can be reliably extracted on the basis of short-time data for pair correlation functions at short spin--spin distance. 
To support this claim further, we compare the post-quench correlation length $\xi_1$ to its thermal equilibrium counterpart.  
As shown in Ref.~\cite{Sachdev1996a.NPhB.464.576S}, the thermal equilibrium correlation length $\xi_T$ can be expressed in terms of a universal scaling function of the temperature $T$. 
Since the one-dimensional TFIM does not have a phase transition at non-zero temperatures, the correlation length remains finite for all non-zero $T$. 
In order to compare $\xi_T$ with the post-quench correlation length $\xi_1$, we need to express $\xi_T$ as a function of $h$ (instead of $T$). The relation between $T$ and $h$ originates from the fact that, upon quenching the ground state of the pre-quench Hamiltonian $H(h_0)$ with the post-quench Hamiltonian $H(h)$, excitations above the ground state of $H(h)$ get populated. In a nonintegrable model, these excitations will thermalize at long times, and the energy content of the excitations will determine the corresponding temperature $T(h)$ in equilibrium. 
In the integrable one-dimensional TFIM, thermalization does not occur in general, and therefore a unique mapping between $h$ and $T$ does not exist.

 \begin{figure}
    \centering
    \includegraphics[width=0.88\columnwidth]{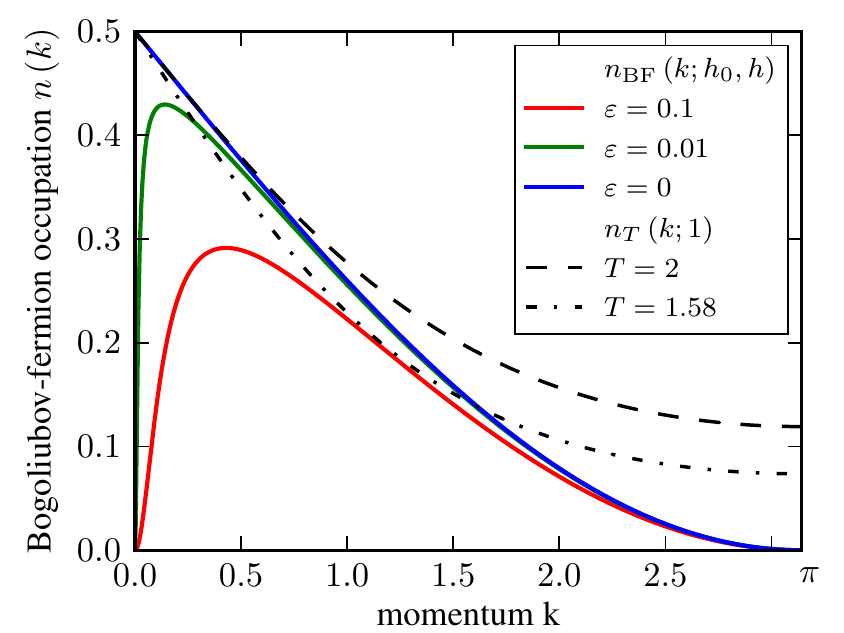}
    \caption{\label{fig:BogOccupations} 
    Occupation numbers of the Bogoliubov fermions as functions of the quasi momentum $k$ after a quench from $h_{0}=10^{3}$ 
    to three different $h=1+\varepsilon$ close to the critical value at $\varepsilon=0$, marked by solid lines.
    The black dashed line marks the thermal occupation number for $h=1$ and an effective temperature $T=2$, cf.~\Eq{Teffective}, while the dotted line is the same distribution at $T=1.58$.
}
\end{figure}

To show this, we consider the occupations \eq{OccNoBogF} of the fermionic Bogoliubov modes diagonalizing the TFIM Hamiltonian after a ground state at $h_{0}$ is suddenly quenched into the vicinity of the critical field strength, $h_{0}>h\gtrsim1$.
These occupations characterize the mapping of the pre-quench ground state onto the eigenmode spectrum of the near-critical Hamiltonian. 
In \Fig{BogOccupations}, they are shown for quenches from $h_{0}=10^{3}$ 
to three different $\varepsilon$ close to the critical value $\varepsilon=0$. 
For comparison, the dashed and dotted lines show thermal occupation numbers 
\begin{equation}
\label{eq:nThermal}
n_{T}(k;h)=\frac{1}{\exp[\omega_{\text{BF}}(k;h)/T]+1}\,,
\end{equation}
for $h=1$, at effective temperatures $T=2$ and $T=1.58$, respectively.
None of the thermal distributions matches the post-quench spectrum over the full range of quasi momenta, which is consistent with the expectation that the steady state reached at long times after the quench is described by a nonthermal generalized Gibbs ensemble.

The low-$k$ mode occupations, however, near $\varepsilon=0$ approach a thermal distribution.
At an effective temperature
\begin{equation}
\label{eq:Teffective}
T = 2\,\frac{h_{0}-1}{h_{0}+1}=\frac{\varepsilon_{0}}{1+\varepsilon_{0}/2}\,,
\end{equation}
and $\varepsilon=0$, the thermal and Bogoliubov-fermion distributions are equal to linear order in $k$, $n_{T}(k;1)=n_{\text{BF}}(k;h_{0},1)+\mathcal{O}(k^{3})$. 
This shows that, when the system is quenched close to $h=1$, the long-wavelength behavior of the fluctuations is equivalent in character to those of a near-critical thermal system.
Note though that at any $h>1$, the zero-mode remains empty since the quench causes modes to become occupied in a pairwise fashion, with Bogoliubov fermion pairs created in modes $k$ and $-k$.

For comparison of the post-quench and thermal correlation lengths, we use Eq.~(A.6) of \cite{Sachdev1996a.NPhB.464.576S}, where the dependence on $h$ of the correlation length for a thermal distribution of the Bogoliubov fermions is made explicit,
\begin{equation}\label{eq:xih}
\xi_{T}^{-1}(h)=2(h-1)\Theta(h-1)-\int_{-\pi}^{\pi}\frac{\mathrm dk}{2\pi}\ln[1-2n_{T}(k;h)]
\end{equation}
with thermal fermion distribution \eq{nThermal}.
We use the temperature $T$ in \eqref{eq:xih} as a free parameter, which we adjust to a value of $1.58$ (in units of $J=1$), chosen such that $\xi_T=\lim_{t\to\infty}\xi_1(t)$ at $\varepsilon=0$ [dashed line in \Fig{correlation_lengths}(c)]. 
Further away from the critical point deviations between $\xi_{T}$ and $\xi_{1}$ are visible, and they may be attributed to two possible reasons: 
First, we fixed $T$ in \eqref{eq:xih} to a specific value, whereas in reality $T$ may effectively vary with $h$, at least for nonintegrable models where thermalization will take place in the long run. 
Second, as discussed above, due to the integrability of the one-dimensional TFIM, the post-quench spectrum corresponds to a generalized Gibbs ensemble  \cite{CalabreseEsslerFagotti11,CalabreseEsslerFagotti12,CalabreseEsslerFagotti12-2}, resulting in deviations from a thermal distribution near the Brillouin edge, cf.~\Fig{BogOccupations}.

Overall these results demonstrate that the short-distance correlation length shows its long-time near-critical behavior already after short evolution times, and that it is consistent with the nondiverging thermal correlation length at $\varepsilon\to0$. 

{\em Scaling.---}%
A further important parallel between post-quench and thermal correlation lengths near the quantum critical point is their potential to exhibit universal scaling behaviour in the vicinity of $h=1$. 
The thermal correlation length at $h=1$ diverges as $\xi_{T}\sim T^{-\nu}$ for $T\to0$, exhibiting the critical exponent $\nu=1$ of the Ising universality class. 

To obtain a diverging post-quench correlation length $\xi_{1}$, the effective temperature \eq{Teffective} needs to be lowered.
For quenches within the paramagnetic regime with $1<h<h_{0}<\infty$, the asymptotic correlation length \eq{xione} is given by
\begin{equation}
  \label{eq:xi1hh0}
  \xi_{1}^{-1}=\ln\left(\frac{2h_{0}h}{h_{0}+h}\right).
\end{equation}
Hence, for $\varepsilon_{0}=h_{0}-1\ll1$ and $h\to1^{+}$ the correlation length scales as $\xi_{1}\sim2/(\varepsilon_{0}+\varepsilon)$, while for $\varepsilon_{0}\gg1$ it approaches the asymptotic value $1/\ln2$ given in \Eq{h0oo}. Critical scaling of the correlation length with respect to the effective temperature scale $T\simeq\varepsilon_{0}$, cf.~\Eq{Teffective}, can therefore be observed when quenching from ground states in the vicinity of the critical point, $h_{0}\gtrsim1$, to $h=1$.
As a consequence, in order to observe scaling behavior in $\varepsilon_{0}$, one needs to tune the pre-quench parameter to different values $h_{0}=1+\varepsilon_{0}\gtrsim1$ close to the critical point, and then to quench to $h=1+\varepsilon\simeq1$.
We point out that it suffices to tune $\varepsilon_{0}$ to a sufficiently small value while the subsequent quench to $\varepsilon$, $0\leq\varepsilon<\varepsilon_{0}$, yields at most a factor of two in $\xi_{1}\simeq2/(\varepsilon_{0}+\varepsilon)$.

To gain intuition on the time evolution of the correlation functions and lengths after such quenches, we repeat the analysis of \Fig{correlation_lengths}(a--c) for quenches with $\varepsilon_{0}\lesssim1$; see \App{ScalingQuenches} for details.
Our main finding, namely that correlation lengths can be extracted at relatively short distances and times, remains valid.
Choosing $\varepsilon_{0}$ closer to the critical point, the correlation length after quenches to $h=1$ increases as $\sim1/\varepsilon_{0}$, and accordingly longer spin chains are required to observe this behavior.

{\em Continuum and scaling limits.---}%
The near-critical scaling  $\xi_{1}\sim\varepsilon_{0}^{-\bar\nu}$ at fixed $\varepsilon=0$, with critical exponent $\bar\nu=1$ is equivalent to the scaling of the TFIM equilibrium correlation length $\xi_{T}\sim T^{-1}$ in the quantum critical regime, i.e., within the ``wedge'' where the temperature is larger than the modulus of the gap,  $T>|\Delta|$.
To show this explicitly we consider the TFIM in the continuum limit, assuming the spins to reside on a lattice with spacing $a$ and taking $a$ to zero. Keeping the sound velocity $c=2Ja$ and the gap energy $\Delta=2|h-J|$ non-zero and finite, it follows that $J$ diverges as $1/a$ and hence the scaling limit must be taken with $\lambda=h/J\to1^{+}$ as $\lambda-1=\Delta/2J=a\Delta/c\to0$. The fermion frequencies reduce to
\begin{equation}
\omega_{\text{BF}}(k,\Delta)=\sqrt{\Delta^{2}+(ck)^{2}}=\Delta_{0}\sqrt{\delta^{2}+\kappa^{2}},
\end{equation}
where we defined
\begin{equation}
\delta=\Delta/\Delta_{0}\qquad\text{and}\qquad\kappa=ck/\Delta_{0}.
\end{equation} 
The gap energy $\Delta_{0}$ (or $\Delta$) now sets the scale of units. 
The mode occupations after a quench from $\Delta_{0}$ to $\Delta$, i.e., from $\delta=1$ to $\delta>0$ in this scaling limit are given by
\begin{equation}
\label{eq:nBFSL}
n_{\text{BF}}(\kappa;\delta)=\frac12\left(1-\frac{\delta+\kappa^{2}}{\sqrt{\delta^{2}+\kappa^{2}}\sqrt{1+\kappa^{2}}}\right).
\end{equation}
Inserting the above expressions and $2J\ln(h_{1})=\sqrt{\Delta\Delta_{0}}$ into \Eq{xione}, the post-quench correlation length $\xi_{1}$ takes on the universal scaling form (cf.~\cite{CalabreseEsslerFagotti12}, Eq.~(271)) 
\begin{align}
 \xi_{1}&=c\Delta_{0}^{-1}F_{1}^{-1}(\Delta/\Delta_{0})\,,
\label{eq:xionesl}
\end{align}
with scaling function (for $x>0$)
\begin{equation}
  F_\text{1}(x)
  =\frac{1+x}{2} + \left(1-\sqrt{x}\right)\Theta(x-1)\,.
  \label{eq:F1}
\end{equation}
The corresponding scaling form of the correlation function,
\begin{align}\label{eq:CzzScaling}
C&^{\sz\sz}(t,r)
 \simeq \left(2\Delta_{0}\right)^{1/4}A_{0}(\Delta/\Delta_{0})\,e^{-|r|/\xi_{1}},
\end{align}
is obtained from \Eq{CzzImproved}, with universal amplitude
\begin{equation}
   A_{0}(x)=\frac{\sqrt{1-x}}{1+x}.
\end{equation}
The thermal correlation length in the scaling limit,
\begin{equation}\label{eq:xiTsl}
  \xi_{T}
  = c\,T^{-1}F^{-1}_\text{T}\left(\Delta/T\right),
\end{equation}
with the universal crossover function
\begin{equation}
  F_\text{T}(x)
  =|x|\Theta(x)+\frac{4\eta}{\pi}\int_{0}^{\infty}dy\ln\coth\frac{\sqrt{x^{2}+y^{2}}}{2}\,,
  \label{eq:FTI}
\end{equation}
was obtained in Ref.~\cite{Sachdev1996a.NPhB.464.576S}, together with the scaling form $C_{T}^{\sz\sz}(r)\sim [T/\sinh(\pi T|r|/c)]^{1/4}$ for the correlation function at temperatures $T\gg\Delta$ near the quantum critical point, reducing to $C_{T}^{\sz\sz}(r)\sim T^{1/4}\exp(-\pi T |r|/4c)$  at $r\to\infty$.
The anomalous exponent $\eta=1/4$ in \eqref{eq:FTI} determines the $r\to0$ spatial equilibrium scaling at the quantum critical point, i.e.,  $C_{T}^{\sz\sz}(r)\sim|r|^{-\eta}$, around $T=0$ and $\Delta=0$.

According to Eqs.~\eq{xionesl} and \eq{F1}, at $\Delta=0$ one has $\xi_{1}=2c/\Delta_{0}$, which coincides with the thermal critical length scale $\xi_{T}(\Delta=0)=4c/(\pi T)$ \cite{Sachdev1996a.NPhB.464.576S} at a temperature $T=T_\text{eff}=2\Delta_{0}/\pi$ defined by the initial-state energy gap. 
Hence, as asserted above, we find the universal scaling $\xi(\Delta=0)\sim1/\Delta_{0}\sim1/T_{\text{eff}}$, which confirms the relation between post-quench and thermal critical behavior close to the quantum critical point.
Note that the post-quench correlation function does not show the thermal algebraic scaling $C_{T}^{\sz\sz}(r)\sim|r|^{-\eta}$ at small $|r|$, consistent with predictions from boundary conformal field theory for the Ising universality class \cite{Calabrese2007a}.

In the scaling window $0\le\Delta\ll\Delta_{0}$ near the critical point, both,  $\xi_{1}^{-1}$ and $\xi_{T_\text{eff}}^{-1}$ deviate from the critical value $\Delta_{0}/(2c)$ by a linear term $\Delta/(2c)$, see the inset of \Fig{correlation_lengths}(d), while only $\xi_{T_\text{eff}}^{-1}$ receives additional corrections, in even powers of $\Delta$ \cite{Sachdev1996a.NPhB.464.576S}.
Figure~\ref{fig:correlation_lengths}(d) in its main part shows the crossover behavior of the post-quench correlation lengths $\xi_{1,2}(\Delta_{0},\Delta)$ as well as of the thermal length $\xi_{T_\text{eff}}(\Delta)$ as functions of the post-quench gap $\Delta$.
All three length scales fall off as $(\Delta/\Delta_{0})^{-1}$ for $\Delta\gg\Delta_{0}$ but there is a factor of two difference between the excited-state post-quench and the zero-temperature equilibrium lengths. 

 \begin{figure}
    \centering
    \includegraphics[width=0.88\columnwidth]{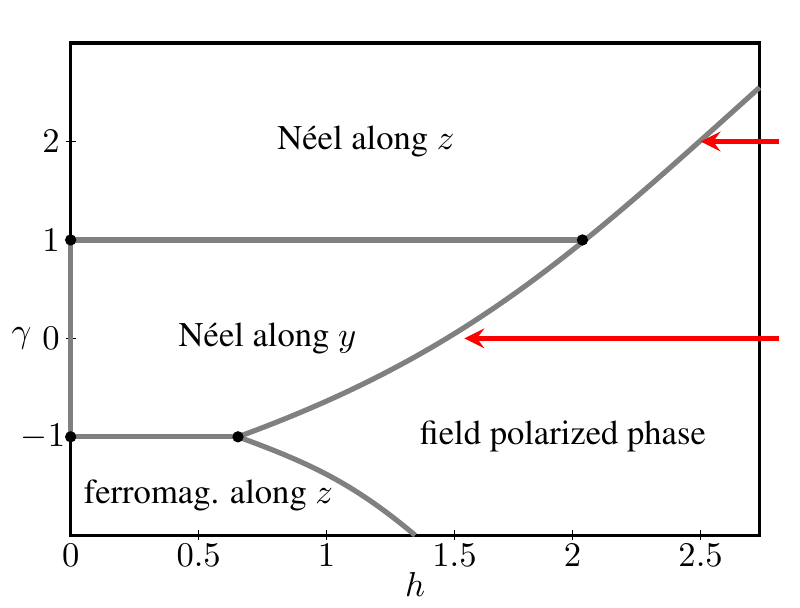}
    \caption{\label{fig:XXZPhaseDiagram} 
    Sketch of the ground-state phase diagram of the 1D Heisenberg XXZ chain, after Ref.~\cite{Dmitriev_etal02}.
    The anisotropy  $\gamma$ and the dimensionless transverse field strength $h$ parametrize the anti-ferromagnetic XXZ Hamiltonian \eq{HXXZ}.
    At $\gamma=1$ one recovers the Heisenberg XXX chain in a transverse field.
    Lines and tricritical points separate the equilibrium phases as indicated.
    We study quenches at fixed $\gamma=0$, $2$, from ground states at large $h$ into the vicinity of the quantum phase transitions at $h\approx1.5$ and $h\approx2.5$, respectively, as marked by the red arrows. 
}
\end{figure}
 \begin{figure*}[ht!]
    \centering
    \includegraphics{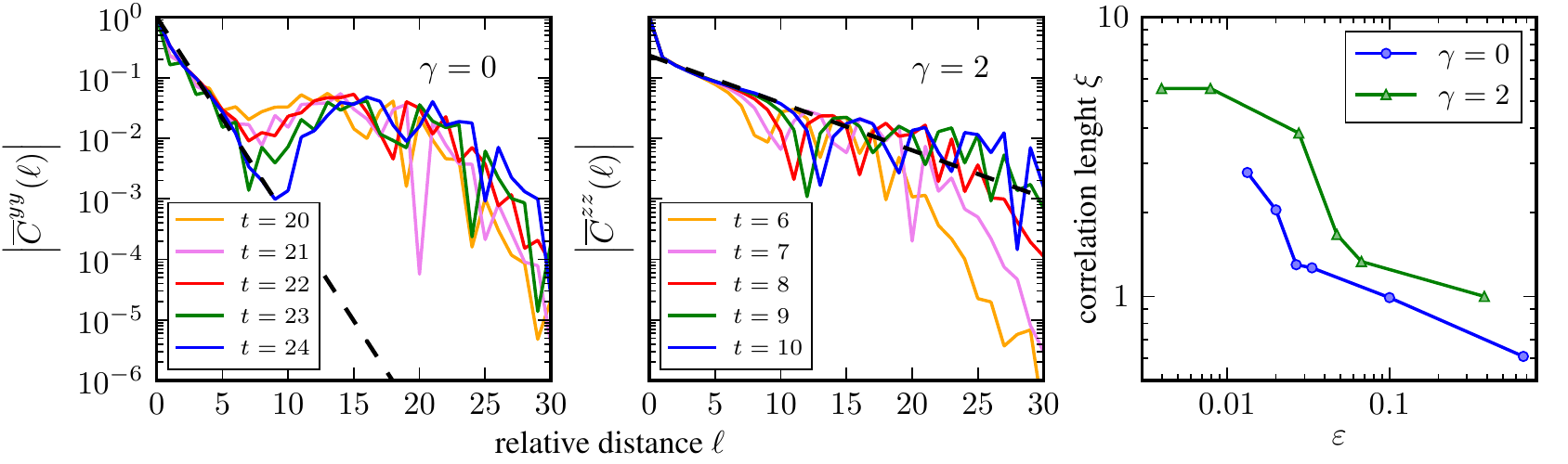}
    \caption{\label{fig:XXZcorrelations} 
    Left and center: Order parameter correlations of the $\mathrm{XXZ}$ chain at various times $t$ after quenches from $h_0\gg1$ into the vicinity of a quantum critical point. The dashed lines are fits to the exponential decays at short distances $\ell$. Data are for anisotropy $\gamma=0$ and post-quench parameter $h=1.54$ (left), and for anisotropy $\gamma=2$ and post-quench parameter $h=2.55$ (center). Right: Correlation lengths $\xi$ extracted by means of exponential fits to the correlation functions. The distances $\varepsilon=h/h_\text{c}-1$ from the critical point are calculated with $h_\text{c}=1.5$ for $\gamma=0$, and $h_\text{c}=2.53$ for $\gamma=2$. The overall shape of $\xi(\varepsilon)$ is consistent with the findings for the one-dimensional TFIM [blue line in \Fig{correlation_lengths}(c)].
}
\end{figure*}

In summary, quenches into the vicinity of the quantum critical point $h=1$ are a viable path for probing thermal-equilibrium critical properties at relatively short times after the quench.
When quenching a ground state at $h>1$, with $\Delta_{0}>0$, to $h=1$, the length scale characterizing the short-distance exponential fall-off of the spatial spin autocorrelation function scales to leading order in $\Delta_{0}$ in the same way as the $h=1$ thermal correlation length in the temperature $T$.
This scheme facilitates the probing of scaling in the quantum critical regime by controlled quenches starting from states outside this regime, without requiring a separate fine tuning of $\Delta_{0}$ and $\Delta$ to criticality.

{\em $\mathrm{XXZ}$ chain.---}To demonstrate the validity of our observations on more general grounds, we consider the one-dimensional anti-ferromagnetic $\mathrm{XXZ}$ chain in a transverse magnetic field,
\begin{equation}
H=\sum_{n=1}^N\left(\sigma_n^\sx\sigma_{n+1}^\sx + \sigma_n^\sy\sigma_{n+1}^\sy + \gamma\,\sigma_n^\sz\sigma_{n+1}^\sz\right) - h\sum_{n=1}^N\sigma_n^\sx,
\label{eq:HXXZ}
\end{equation}
where $\gamma\in\RR$ is the anisotropy parameter and we have set the Heisenberg coupling to $J=-1$. 
For any transverse-field strength $h$, the model is known to be integrable in the isotropic case $\gamma=1$ as well as for $\gamma\to\pm\infty$ \cite{Dmitriev_etal02}, but nonintegrable otherwise. 
This allows us to tune the model away from integrability by varying $\gamma$, while using $h$ as an independent quench parameter. 
The ground state phase diagram of the model, comprising four different phases, is fairly well-known, see \cite{Dmitriev_etal02} and \Fig{XXZPhaseDiagram}. 
We choose the two cases $\gamma=0$ and $\gamma=2$, both of which are deep in the nonintegrable regime. 
As in the protocol we used for the TFIM, we start in the paramagnetic phase from the fully $x$-polarized ground state at $h\gg1$. 
In the case $\gamma=0$ we quench into the vicinity of the quantum critical point at around $h\approx1.5$, which separates the paramagnetic phase from a phase with N{\'e}el ordering in the $y$ direction. 
In the case $\gamma=2$ we quench to the quantum critical point around $h\approx2.5$, which separates the paramagnetic phase from a phase with anti-ferromagnetic order along the $z$ direction; see Fig.~1 of \cite{Dmitriev_etal02}. 

We use the infinite-system density matrix renormalization group (iDMRG) \cite{McCulloch07,McCulloch,MichelMcCulloch,Hubig_etal15,Hubig_etal17} based on matrix product states (MPS) \cite{McCulloch07,Schollwoeck11}, and carry out the time-evolution simulations using time-evolving block decimation \cite{Vidal03,Verstraete_etal04,ZwolakVidal04}. In the framework of iDMRG, one uses a truncation threshold \cite{White92,White93,Schollwoeck05}, which sets an upper bound on the norm distance between the exactly-evolved state and its approximated counterpart in the simulation. 
We use a truncation threshold of $10^{-12}$ per time step, with a time step of $\Delta t = 0.01$. Correspondingly, we do not set an upper limit on the MPS bond dimension, thereby allowing the latter to grow freely (exponentially) in evolution time in order to always achieve the truncation threshold we set. This eventually exceeds the available numerical resources, which in turn limits the range of accessible evolution times. Nevertheless, even with our stringent accuracy goals, we reach sufficiently long evolution times for our purposes.

Fig.~\ref{fig:XXZcorrelations} shows the thus obtained equal-time correlations $C^{\sy\sy}$ (left) and $C^{\sz\sz}$ (center) as a function of the spin--spin separation $\ell$ for various $t$. To suppress fluctuations, the plots show correlations averaged over short intervals of time,
\begin{equation}
\overline{C}^{aa}(t,\ell):=\frac{1}{\tau}\int_{t-\tau}^t dt' C^{aa}(t',\ell),
\end{equation}
where $a$ denotes any of the spin components. We chose $\tau=1$ for the averaging interval. At times around $t=10$ or 20 (depending on the model parameters), averaged correlations are found to have essentially converged to an exponential fall-off that extends over ten or more lattice sites. The corresponding correlation length $\xi$, obtained by fitting an exponential function to the fall-off, is shown in Fig.~\ref{fig:XXZcorrelations} (right). Qualitatively the behavior of $\xi$ as a function of the distance $\varepsilon$ from the critical point is similar to the analytical results for the one-dimensional TFIM reported in \Fig{correlation_lengths}(c).

{\em Conclusions.---}In summary, we have demonstrated that equilibrium near-critical properties can be extracted from correlation functions at short times and short distances after quenches into the vicinity of a quantum critical point. 
Our results were obtained for one-dimensional spin models, either integrable or nonintegrable, so that either analytic calculations or $t$DMRG calculations are feasible, but we do not see a reason why the same kind of quench protocol should not be applicable also in higher dimensions. 
The time scales after which equilibrium properties can be extracted appear sufficiently short compared to coherence time scales of experimental quantum simulators of spin models, and determining equilibrium scaling laws should therefore be feasible with these platforms. 
We believe that a crucial ingredient that leads to the short equilibration time scales after the quench is the fact that the initial state, which is chosen such that it can easily be prepared experimentally, is spatially homogeneous, and not too different from the (likewise spatially homogeneous) equilibrium state after the quench. 
As a consequence, local equilibration, which happens fast, appears to imply also global equilibration.

We finally remark that, for classical dissipative as well as quantum systems, various short to intermediate-time scaling phenomena such as initial-slip scaling and critical ageing \cite{Janssen1989a,Janssen1992a,Diehl:1993az,Diehl1993b,Li1995a.PhysRevLett.74.3396,Okano1997a,Calabrese2002a.PhysRevE.65.066120,Calabrese2005a.JPA38.05.R133,Gambassi2006a.JPAConfSer.40.2006.13}, as well as coarsening dynamics \cite{Bray1991a.PhysRevLett.67.2670,Bray1994a.AdvPhys.43.357,Bray2000PhRvL..84.1503B,Mukerjee2007a.PhysRevB.76.104519,Williamson2016a.PhysRevLett.116.025301,Hofmann2014PhRvL.113i5702H,Williamson2016a.PhysRevA.94.023608,Bourges2016a.arXiv161108922B.PhysRevA.95.023616} and the approach of a non-thermal fixed point  \cite{Berges:2008wm,Berges:2008sr,Scheppach:2009wu,Nowak:2011sk,Schole:2012kt,Nowak:2012gd,Karl:2013mn,Karl:2013kua,Orioli:2015dxa} are known.
In these systems,  long-wavelength modes are frequently observed to equilibrate slowly, particularly in more than one spatial dimension.
Similarly, for the systems studied in the present paper, equilibration of the long-distance correlations is found to be particularly slow, and for the TFIM the length scale on which correlations are equilibrated grows only logarithmically with time. Likewise, the fast equilibration of short-distance correlations we observe is reminiscent of the rapid thermalization of short-wavelength modes in many of the above cited examples of universal scaling dynamics.

We thank Ian P.~McCulloch for providing the iDMRG toolkit used for the
simulations in this work \footnote{See \url{https://people.smp.uq.edu.au/IanMcCulloch/mptoolkit/}}. Collaboration and discussions with J.~Berges, S.~Czischek, S.~Diehl, S.~Erne, F.~Ess\-ler, M.~Fagotti, M.~G\"arttner, P.~Kunkel, D.~Linnemann, J.~Marino, W.~Muessel, J.~M.~Pawlowski, M.~Pr\"ufer, M.~Rabel, C.~Schmied, and H.~Strobel are gratefully acknowledged.
T.~G., M.~Karl, and M.~K.~O.~acknowledge funding by the EU (ERC advanced grant EntangleGen, Project No.~694561, FET-Proactive grant AQuS, Project No.~640800).
M.~Kastner acknowledges kind hospitality at the Kirchhoff-Institut f\"ur Physik and financial support from the Fakult\"at f\"ur Physik at the Ruprecht-Karls-Universit\"at Heidelberg, where part of this work was done; as well as financial support by the National Research Foundation of South Africa through the Incentive Funding and the Competitive Programme for Rated Researchers.
\\


\appheading{}
\appendix

\section{Long-range correlations at finite times}
\label{app:LongRange}
%
 \begin{figure}
    \centering
    \includegraphics[width=\columnwidth]{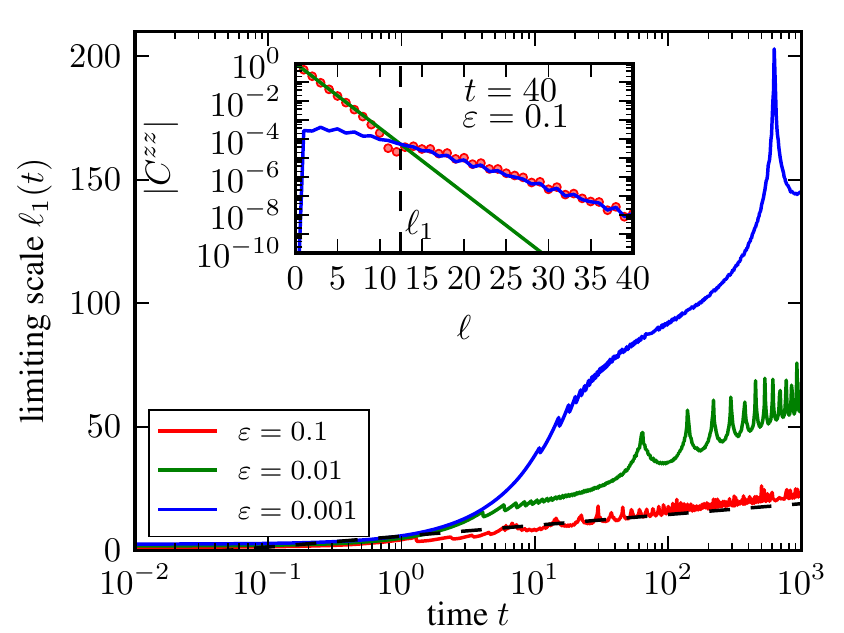}
    \caption{\label{fig:ScaleEll1} 
    Time dependence of the limiting scale $\ell_{1}(t)$ for three different quenches, from a large value of $h_{0}=10^3$ to $\varepsilon=0.1$ (red), $\varepsilon=0.01$ (green), and $\varepsilon=0.001$ (blue), on a semi-logarithmic scale.
    $\ell_{1}(t)$ is extracted as the distance $\ell$ where the length scales defined by the two terms in \Eq{CzzImproved}, the asymptotic exponential $\sim\exp\{-\ell/\xi_{1}\}$ and the transient integral contribution, intersect; see inset. 
    The dashed line marks the analytically determined growth $\ell_{1}(t)\sim (3\xi_{1}/2)\ln(t/t_{0})$ (cf.~Eq.~(37) of Ref.~\cite{CalabreseEsslerFagotti12}) for quenches to $\varepsilon\gtrsim0.1$.
    After an initial period where the exponential correlations build up, the transition scale $\ell_{1}$ oscillates, with envelope increasing logarithmically in time, the slower, the smaller $\xi_{1}$ is.  
}
\end{figure}
\begin{figure*}[th!]
  \setcounter{section}{2}
  \setcounter{figure}{0}
    \centering
    \includegraphics[]{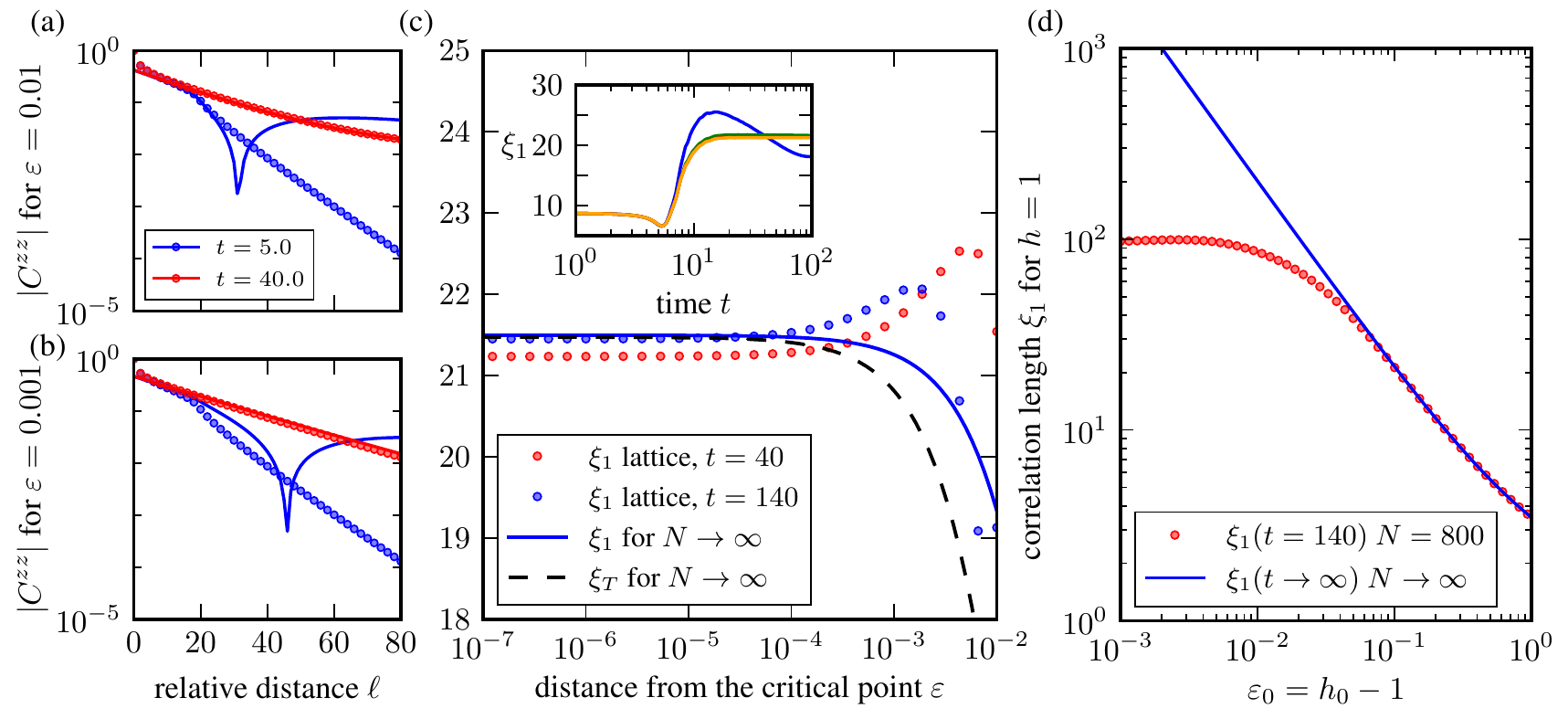}
    \caption{\label{fig:correlation_lengths_scalingregime} 
   Order-parameter correlation functions and correlation lengths after a quench from $h_{0}=1.1$ into the vicinity of the quantum phase transition of the one-dimensional transverse-field Ising model.
   (a) Correlation function $|C^{\sz\sz}|$, on a log scale as a function of the spin--spin separation $\ell$ at two different times $t$ after the quench from $h_{0}=1.1$ to $\varepsilon=h-1=0.01$. 
   Circles mark the exact solution obtained by diagonalisation in terms of Bogoliubov fermions, while the solid line shows the approximate function \eq{CzzImproved}.
   (b) As in (a), but for a quench to $\varepsilon=0.001$.
   (c) Inset: Time evolution of the correlation length $\xi_{1}$, obtained by fitting exponential functions to the short-distance ($\ell\in\{21,...,30\}$) fall-off of $|C^{\sz\sz}|$ for finite spin chains. Data for four different quenches are shown, with final quench parameters $\varepsilon=0.01$ (blue), $0.001$ (green), $0.0001$ (red), and $0.00001$ (yellow). 
   Main graph: 
   Correlation length $\xi_{1}$ as a function of the post-quench parameter $\varepsilon$. Circles 
   are obtained from exponential fits to the data for $\ell\in\{21,...,30\}$, at times $t=40$ (red) and $t=140$ (blue circles).
   The blue line is the analytical expression \eqref{eq:h0oo} for $\xi_{1}$ in the thermodynamic limit. 
   $\xi_{T}$ marks the thermal correlation length \eq{xih} for an effective temperature $T=0.12$.
   The total number of spins in the periodic chain in the numerical computations is $N=800$, which is large enough to avoid finite-size effects.
   (d) Correlation length $\xi_{1}$ for quenches from different $h_{0}$ to $h=1$ for a chain of $N=800$ spins at time $t=140$ after the quench (red circles).
   The blue solid line depicts the asymptotic behavior ($t\to\infty$) in the thermodynamic limit.
}
\end{figure*}
 \begin{figure}[h!]
    \centering
    \includegraphics[width=0.88\columnwidth]{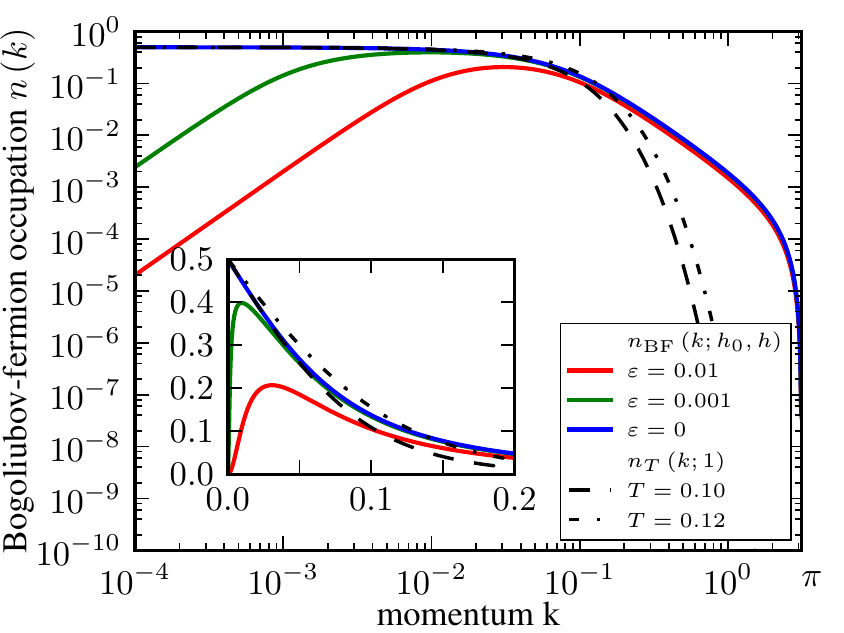}
    \caption{\label{fig:BogOccupationsScalReg} 
    Occupation numbers of the Bogoliubov fermions as functions of the quasi momentum $k$ after a quench from $h_{0}=1.1$ to three different $h=1+\varepsilon$ close to the critical value at $\varepsilon=0$, marked by solid lines.
    The black dashed line marks the thermal occupation number for $h=1$ and an effective temperature $T=0.1\simeq2(h_{0}-1)/(h_{0}+1)$, while the dotted line is the same distribution at $T=0.12$, leading to the same $\varepsilon\to0$ limit of the post-quench and thermal correlation length, as seen in \Fig{correlation_lengths_scalingregime}(b).
}
  \setcounter{section}{1}
  \setcounter{figure}{1}
\end{figure}

Here, we discuss the transient long-range spatial correlations appearing after the quench applied to the initial ground state of the TFIM Hamiltonian. 
In \Figs{correlation_lengths}(a) and (b) we show snapshots of the time evolution of the correlation function for quenches to $\varepsilon=0.1$ and $\varepsilon=0.001$, at times $t=5.0$ and $t=40.0$. 
At short distances $\ell\lesssim\ell_{1}(t)$, below a limiting length scale $\ell_{1}$, the asymptotic long-time behavior $C^{\sz\sz}(t,\ell)\simeq \exp(-\ell/\xi_{1})$ is already attained to a very good degree. 
The weaker decay at larger distances is described by the second, integral term in \Eq{CzzImproved}.
Moreover, in accordance with Eq.~(37) in \cite{CalabreseEsslerFagotti12}, the limiting scale $\ell_{1}$ increases logarithmically in time.
$\ell_{1}(t)$ can be defined as the crossing point of the first and second terms in \eq{CzzImproved}, and is shown in \Fig{ScaleEll1}.
For quenches to $\varepsilon\gtrsim0.1$, the scale increases as $\ell_{1}(t)\sim (3\xi_{1}/2)\ln(t/t_{0})$. 
Beyond $\ell_{1}$, the correlations show a spatial decay with superimposed oscillations, with an envelope decaying proportional to $\exp(-\ell/\xi_{2})$ with a second characteristic length scale
\begin{align}\label{eq:xitwo}
    \xi_2^{-1} &= -\frac{1}{2\pi} \int_{-\pi}^{\pi}\mathrm dk \, \ln\left|1 - 2 n_{\text{BF}}(k)\right|. 
\end{align}
In the limit $h_{0}\to\infty$, the correlation length \eq{xitwo} simplifies to
\begin{equation}\label{eq:xi2h0oo}
 \xi_{2}^{-1}=\ln(2h)-\Theta(h-1)\mathrm{arcosh}(h),
\end{equation}
shown as a green line in \Fig{correlation_lengths}(c), respectively.
In the limit $\varepsilon=h-1\to0^{+}$, the two correlation lengths equal each other, $\xi_{2}\to\xi_{1}+0^{+}$. 

In the continuum and scaling limits, inserting the expression \eq{nBFSL}  for the occupation number distribution $n_{\text BF}$ and $2J\ln(h_{1})=\sqrt{\Delta\Delta_{0}}$ into \Eq{xitwo}, the correlation length $\xi_{2}$ takes on the form (cf.~\cite{CalabreseEsslerFagotti12}, Eq.~(271))
\begin{align}
  \xi_{2}&=c\Delta_{0}^{-1}F_{2}^{-1}(\Delta/\Delta_{0})\,,
  \label{eq:xionesl2}
\end{align}
with universal scaling function
\begin{align}
  F_{2}(x)&
  =\frac{1+x}{2}-\sqrt{x}.
  \label{eq:F2}
\end{align}
This is shown as a green line in \Fig{correlation_lengths}(d).

\section{Quenches in the vicinity of the critical point}
\label{app:ScalingQuenches}
  \setcounter{section}{0}
  \setcounter{figure}{0}

Here we provide numerical results for quenches in the scaling regime, from $h_{0}\lesssim2$ to $1< h\ll2$.
In \Figs{correlation_lengths_scalingregime}(a) and (b), we compare exact results for the correlation function $C^{\sz\sz}$ to the asymptotic formula \eq{CzzImproved}, finding good agreement at rather early times $t<40$. 
At short distances $10\lesssim\ell\lesssim 60$, the asymptotic long-time behavior $C^{\sz\sz}(t,\ell)\simeq \exp(-\ell/\xi_{1})$ is already attained to a very good degree. 

We extract the correlation length $\xi_{1}$ from the data for the correlation functions shown in \Figs{correlation_lengths_scalingregime}(a, b) by fitting exponential functions to the short-distance decay of $C^{\sz\sz}$.
When doing so, we discard the short-range fall-off near $\ell=0$, which is due to the finite size of the system, by fitting to the data in the range   $\ell\in\{11,...,20\}$.
The time-dependence of $\xi_{1}$ is shown in \Fig{correlation_lengths_scalingregime}(c) (inset) for various $\varepsilon$, and stationarity is observed after times $t\gtrsim10$ for small $\varepsilon$.

As \Fig{BogOccupations} does for quenches from large $h_{0}$, \Fig{BogOccupationsScalReg} shows the occupation numbers \eq{OccNoBogF} of the Bogoliubov fermions after a quench from $h_{0}=1.1$ into the vicinity of the quantum critical point.
The colored lines mark different values of the post-quench parameter $\varepsilon$ close to the critical value $\varepsilon=0$. The dashed and dotted lines show the thermal occupation numbers at effective temperatures $T=0.1$, approximating $T=2(h_{0}-1)/(h_{0}+1)$ as obtained in \Eq{Teffective}, and $T=0.12$ for which the thermal and post-quench correlation lengths are equal at $\varepsilon\to0^{+}$.
None of the thermal distributions matches the post-quench spectrum over the full range of quasi momenta, which is consistent with the expectation that the steady state reached at long times after the quench is described by a nonthermal generalized Gibbs ensemble.

In the thermodynamic limit $N\to\infty$, the correlation length \eq{xione} simplifies to the expression given in \Eq{xi1hh0}, which in the main graph of \Fig{correlation_lengths_scalingregime}(c) is compared with the thermal correlation length $\xi_{T}$, \Eq{xih}, for an effective temperature $T=0.12$.

The asymptotic post-quench correlation length $\xi_{1}$ is, furthermore, compared with its values at times $t=40$ (red circles) and $t=140$ (blue circles) after the quench. 
The plot illustrates that for quenches within the scaling regime the approach to the asymptotic correlation length takes longer than for the quenches considered in \Fig{correlation_lengths}.
\Fig{correlation_lengths_scalingregime}(d) demonstrates the scaling of $\xi_{1}\sim\varepsilon_{0}^{-1}$ for quenches from $h_{0}<1$ to $h=1$.
The saturation at small $\varepsilon_{0}=h_{0}-1$ is due to the finite time after the quench and the finite length of the chain.

We finally consider the time scale within which the asymptotic correlation length is reached in the scaling region.
Within an equilibrium approach to the TFIM one finds that the relaxation time $\tau_{T}$, in the quantum critical region ($T\to0$, $T\gg|\Delta|$), is related to the correlation length by $\tau_{T}=\xi_{T}\pi\cot(\pi/16)/(8c)\simeq1.98\,\xi_{T}/c$ \cite{Sachdev1996a.NPhB.464.576S}.
At the time $t=140$ shown in \Fig{correlation_lengths_scalingregime}(d), the correlation length has reached its asymptotic value already down to $\varepsilon\simeq0.06$ where it takes the value $\xi_{1}(t=140;\varepsilon_{0}=0.06)\simeq33$.
Assuming that the same relation as in equilibrium holds for the relation between $\xi_{1}$ and a corresponding asymptotic relaxation scale $\tau_{1}$, one finds  $\tau_{1}\simeq65\simeq0.5\,t$ which is consistent, by order of magnitude with the time required for reaching the asymptotic limit.


%

\end{document}